# Single-Crystal Silicon Thermoelectrics by Phonon Engineering

Thierno-Moussa Bah[1, 2], Stanislav Didenko[2], Di Zhou[2], Tianqi Zhu[2], Hafsa Ikzibane[2], Stephane Monfray[1], Thomas Skotnicki[3], Emmanuel Dubois[2], Jean-François Robillard[2]

[1] STMicroelectronics - 850 rue jean Monnet 38920 Crolles (France),
[2] Univ. Lille, CNRS, Centrale Lille, Junia, Univ. Polytechnique Hauts-de-France, UMR 8520 - IEMN – Institut d'Electronique de Microélectronique et de Nanotechnologie, F-59000 Lille, France
[3] CEZAMAT, Warszawa, Poland
E-mail : jean-francois.robillard@iemn.fr

## Abstract

Herein, we report the use of nanostructured crystalline Si as a thermoelectric material and its integration into thermoelectric harvesters. The proof-of-concept relies on the partial suppression of lattice thermal transport by introducing pores with dimensions scaling between the electron mean free path and the phonon mean free path. In other words, we artificially aimed at the electron crystal–phonon glass trade-off targeted for thermoelectric efficiency. The devices were fabricated using CMOS-compatible processes and exhibited power generation from a few μW/cm² to a few mW/cm² under temperature differences from a few K to 200 K across the thermopiles. These numbers demonstrate the capability to power autonomous devices with environmental or body heat using silicon chips with areas below cm². This paper also reports the possibility of using the developed demonstrators for integrated thermoelectric cooling.

## Introduction

The blooming of the Internet of Things (IoT) and wireless autonomous sensor nodes has been delayed owing to the lack of reliable, safe, and low-cost energy sources [1]. Thermoelectric generators (TEG) have these advantages [2]. Silicon has the advantages of being abundant, non-toxic, and abundant facilities and technological processes for low-cost mass production compared to conventional thermoelectric materials (bismuth telluride alloys). The development of Si-based TEG [3–5] has attracted increasing research interest. However, silicon is an inefficient thermoelectric material owing to its high thermal conductivity of 150 W/m/K [6], whereas bismuth telluride alloys exhibit hundred times lower conductivity (~1.5 W/m/K) [7]. Moreover, the dimensionless figure of merit $zT$ which characterizes the thermoelectric efficiency, is approximately 1 for $Bi_2Te_3$ at room temperature [8] and below 0.01 for bulk silicon [9]. At temperature $T$, $zT$ is defined as

$$zT = \frac{\sigma \cdot S^2}{\kappa} \cdot T \qquad (1)$$

$\sigma$, $S$ and $\kappa$, are the electrical conductivity, Seebeck coefficient, and thermal conductivity, respectively. The thermal conductivity accounts for the performance gap between the Si and $Bi_2Te_3$ micro-TEGs (Figure 1). Efforts are oriented towards cutting the phonon part of heat transport, which is the dominant contribution in semiconductors [10]. The phonon contribution is evaluated in the diffusive regime as

$$\kappa_{ph} = \frac{c \cdot v \cdot \Lambda}{3} \qquad (2)$$

where c, $v$ and Λ are the specific heat, group velocity (average elastic wave velocity), and phonon mean free path, respectively.

The heat capacity depends on the atomic density of the material, and thus cannot be modified. Group velocity modulation would require modifying the atomic structure of the crystal to induce coherent effects, unexpected at room temperature. Suppression of the lattice contribution can be achieved by reducing the phonon mean free path through diffusion through artificial defects and boundaries. This process can be achieved by the use of thin films [6], nanowires [11], the addition of impurities or polycrystalline silicon [12], and silicon nanopatterning [13,14]. Previous studies have shown that holey structures can further downscale the thermal conductivity with a minor impact on the electrical conductivity [15] in a so-called Phonon Engineering (PE) approach. The combination of thin films and PE lowers the thermal conductivity to 34 W/m/K [13] and even to 2 W/m/K [14,15]. The first value, obtained from our own measurements, was for samples with partially perforated patterns. This constitutes a conservative hypothesis. The lowest value represents the amorphous limit of silicon thermal conductivity [16], which makes it a very ambitious limit that can probably be obtained only at the cost of substantial material defects.

In addition to reducing the thermal conductivity of silicon, research in recent years has dealt with the development of silicon-based (mainly polysilicon and silicon nanowires) thermoelectric micro-harvesters or generators. Figure 1 reports the main results [3,5,17–20] from state-of-the-art silicon-based micro-harvesters with respect to state-of-the-art bismuth telluride harvesters [21]. The state-of-the-art silicon TEGs exhibits barely a few µW/cm$^{-2}$ against some mW/cm$^{-2}$ for vertical Bi$_2$Te$_3$ [21,22] TEG at the same temperature differences. The thermoelectric properties of these materials primarily explain this gap.

In this study, we demonstrate a proof-of-concept thermoelectric harvester using PE silicon membranes. In the following section, we present the harvester fabrication process and discuss its performance with respect to a non-phonon-engineered prototype, especially according to the state-of-the-art performances presented in Figure 1. This paper also insists on the modeling results reported in [23] and stipulates that the performance of PE Si-based TEG compares favorably with that of Bi$_2$Te$_3$ based TEG when the heat flow is not controlled by a bulky heat sink (*i.e.*, in the specific use case for IoT, which is deeply constrained by compactness and low form factor requirements).

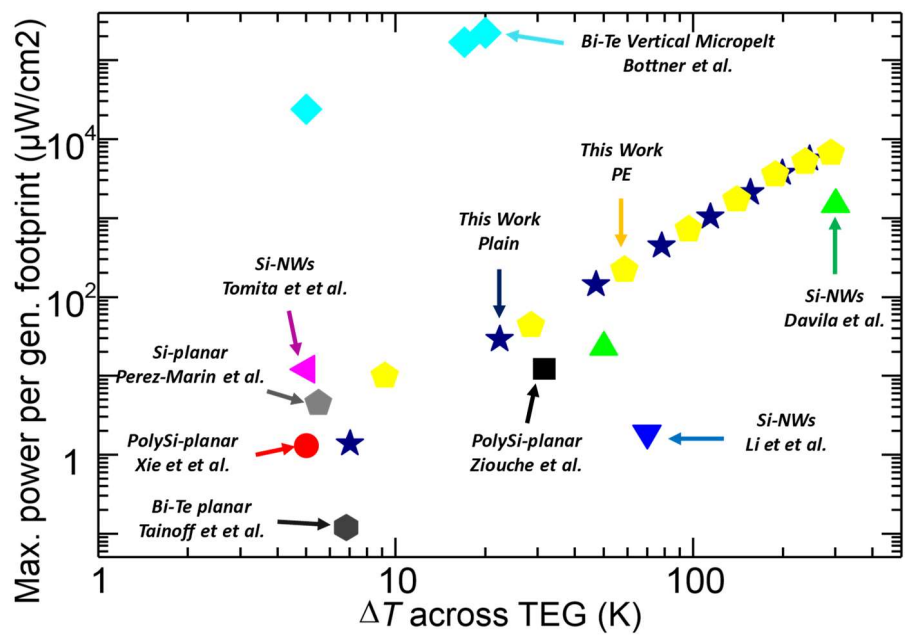

## Experimental Methods

The demonstrators were fabricated from an SOI wafer using the conventional CMOS processes (Figure 2). The process is mainly based on our previous work carried out to reduce the thermal conductivity of silicon by means of silicon thinning coupled to PE [13,25]. The departure point was an SOI wafer with a 70 nm thick active layer (a) upon high-resolution e-beam lithography, and $Cl_2$-Ar reactive ion etching (RIE) was performed to define the PE patterns (b). Then, the silicon is *p*-and *n* doped at $6 \cdot 10^{18}$ cm$^{-3}$ and $9 \cdot 10^{18}$ cm$^{-3}$ respectively, by ion implantation to create thermopiles (c). A low-stress silicon nitride was then deposited (d) to insulate the thermopiles from a platinum resistive heater, aiming to emulate the demonstrators' hot sources using Joule heating. Silicon nitride also aims to improve the mechanical strength of structures. Cavities are etched around the thermopiles as apertures for subsequent suspension (e). Metallic sensors and leads are obtained by platinum and gold evaporation (f).

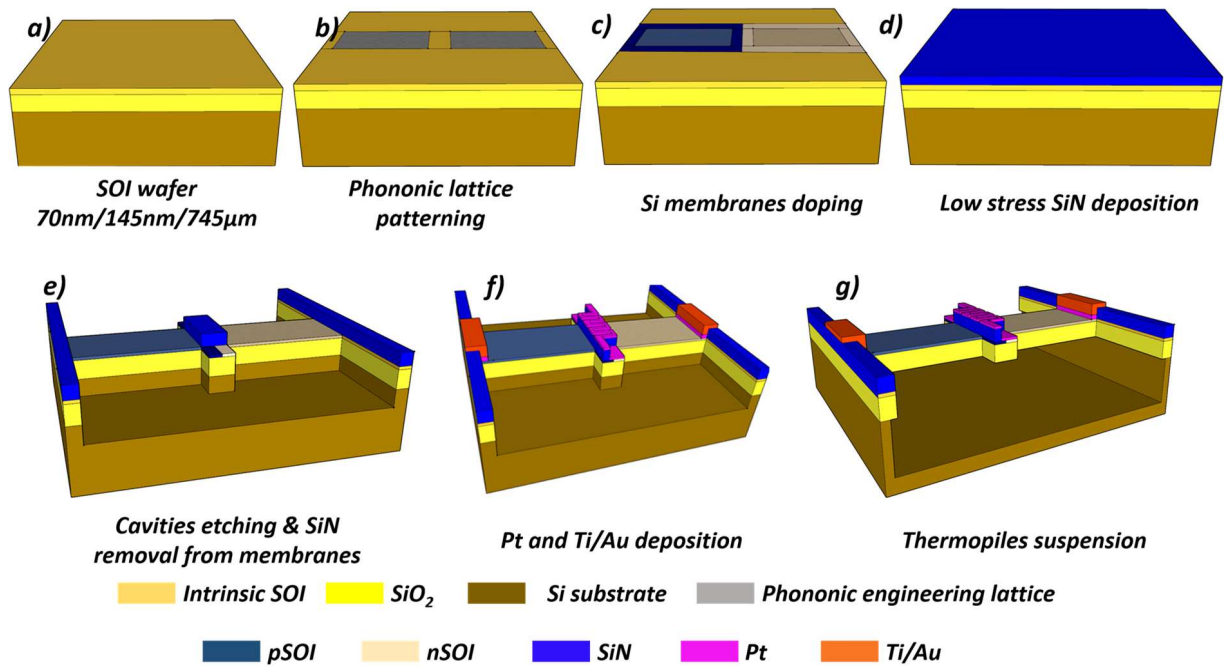

Before metallization, the Si membranes were coated with thermal oxide to protect them from xenon-difluoride (XeF$_2$) etching (g). This last step finishes the process and ensures thermal insulation of the Si membranes from the wafer. The device was made of five thermopiles connected electrically in series and thermally in parallel (Figure 3-a). It was equipped with four pads to supply heating power and I-V measurements. Each thermopile (50µm long, 10µm wide and 205 nm thick) was composed of two *n*- and *p*-doped membranes, as shown in Figure. 3-b. The membrane stack is composed of the following components:

15nm SiO$_2$/62 nm Si/ 145 nm SiO$_2$. At the center, two platinum islands electrically short-circuited the p-n boundary (Figure 3-b). Platinum resistive heaters were deposited over the silicon nitride layer to avoid current leakage (Figure 3-b, c, d). A PE lattice with 40 nm diameter pores and 100 nm pitch patterns was used for each silicon membrane (Figure 3-e, f). The electrical continuity of the thermopiles was ensured using a metallic Pt strap between adjacent thermopiles.

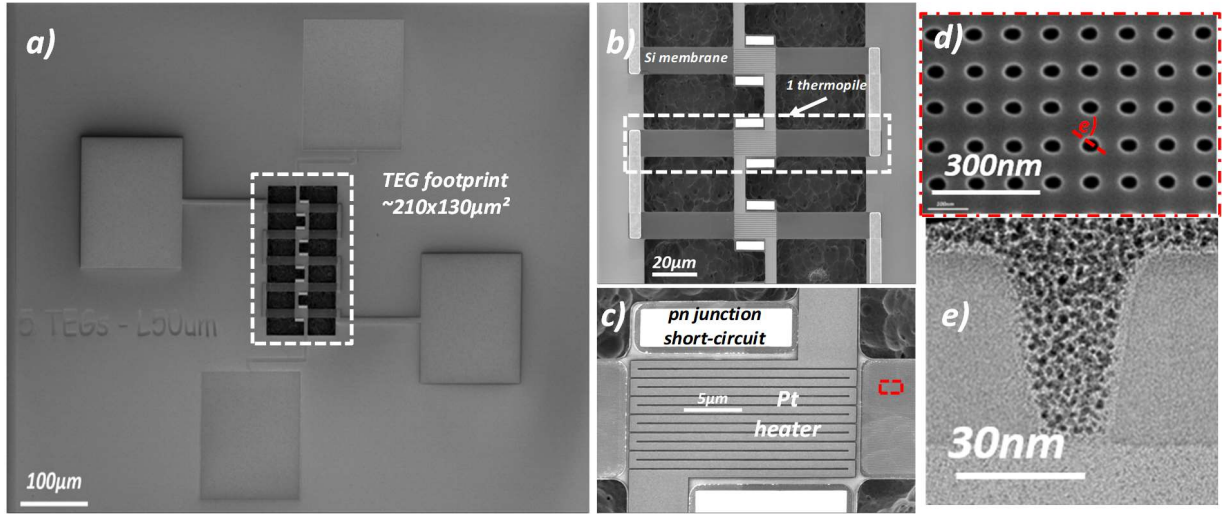

***Figure 3:*** *(a) Scanning Electron Micrograph of a thermoelectric generator demonstrator made of 5 suspended thermopiles. (b) Close-up view of three thermopiles. (c) Platinum resistive heater. (d) Detail of the membranes PE patterns. (e)Cross section Transmission Electron Micrograph of the membrane*

The demonstrators were studied both in terms of energy production under a temperature gradient (Seebeck effect) and thermoelectric cooling when current flows across thermopiles (Peltier effect). Regarding Seebeck effect, the devices were characterized by a thermo-resistive methodology [14,26,27]. Indeed, Pt is known to have a significant temperature-dependent coefficient of resistance α=2.5×10$^{-3}$/K in this case. This value differs from the bulk value because the Pt film is 30 nm thin and was systematically measured for each sample before the thermometry measurement. Infrared imaging (Figure 4-b) demonstrates the functionality of these heaters. The electrical power ($P_H$) produced at the center of the thermopiles is converted into heat flux ($Q_H$) and, in the first approximation, equally diffuses across the thermopiles (Figure 4-c) to the cold ends. Measurements were performed under vacuum to eliminate conducto-convection in air and guide heat flux into the membranes. During the measurements, the frame wafer was maintained at a constant temperature of 25°C. Because the thermopiles were etched from the SOI substrate, efficient heat coupling was achieved at the cold ends. This assumption is supported by the high thermal conductance of the frame wafer and the infrared thermometry. It follows that the cold sides of the membranes are at the same temperature as the substrate, which was verified by IR imaging. The characterization was performed in two steps: i) by variation of the Pt heating current, a V–ΔT characteristic enables measurement of the Seebeck coefficient, and ii) at a constant Pt heating current, the I-V curves characterize the power generation. To capture the Peltier effect, a current was applied to the thermopiles and imaged using IR thermometry in ambient air. This electrical current should produce heat absorption or generation at the center of the thermopiles depending on its direction. IR imaging requires heating the samples to 75°C during acquisition to maximize accuracy.

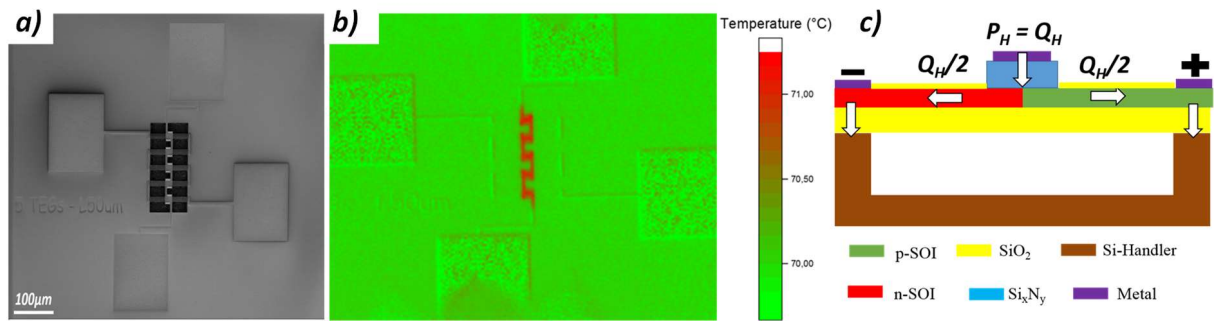

*Figure 4: SEM view of a 5 thermopiles demonstrator (a) IR imaging (performed in ambient air) after Pt resistive serpentines voltage bias (b) and cross-sectional view of a thermopile with the heat diffusion across it (c)*

### Results & Discussions

Figure 5 shows the temperature difference across the thermopiles as a function of heating power for the two plain and two PE devices. The PE thermopiles sustained a higher temperature difference for a given amount of heating power. Thus, the thermal resistance of the thermoelectric leg increased with respect to the plain case. There is a two-fold interest in this result. First, the use of a thin-film planar device significantly increased the thermal resistance to approximately $10^6$ K/W. This value governs the temperature difference in a TEG and should be compared to the typical BiTe TE legs of $10^4$ K/W [22] ($10^2$ K/W for bulk Si with typical leg dimensions of $1\text{mm} \times 1\text{mm} \times 5\text{mm}$). In addition to the thin-film geometry, PE enhances thermal resistance by a factor of 1.8. The reduction in thermal conductivity reduction [13,14] with PE explains this gain.

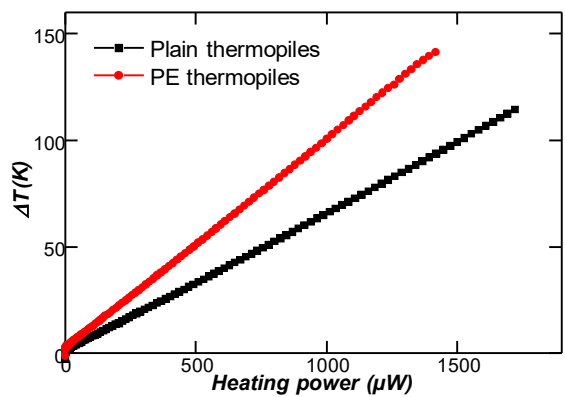

**Figure 5:** *Pt serpentines electrical resistances variation of a 5 thermopiles device as a function of the heating power.*

Figure 6 shows the Seebeck voltage generated as a function of temperature difference across the thermopiles. Again, characterization was performed on plain and PE samples. As expected, the Seebeck voltage increased linearly with the temperature gradient. Furthermore, the plain demonstrators produced an electrical voltage close to the value expected by a Finite Element Method (FEM) software using a Seebeck coefficient $S_p\text{-}S_n$= 545μV K$^{-1}$. The FEM cannot model the physical effect of PE; therefore, the modeling results can only be compared to those of plain demonstrators. The measurements showed Seebeck coefficients per thermopile of 570μV/K and 590μV/K for the plain demonstrators and 820μV/K and 840μV/K for the PE demonstrators. There was a 1.4 times increase induced by PE.

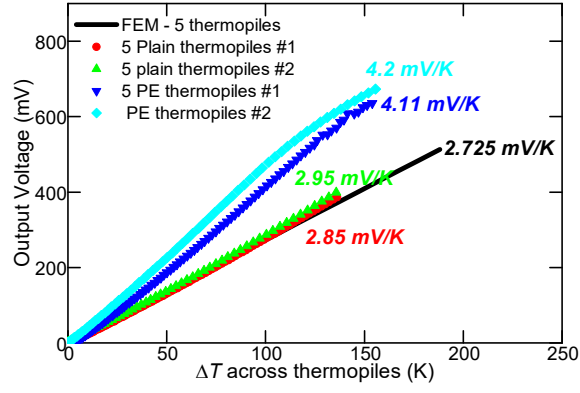

**Figure 6:** *5 plain thermopiles (black squares & red dots) and 5 PE thermopiles (green & blue triangles) output voltage with respect to the temperature difference across thermopiles*

Power generation was investigated by studying the I-V characteristics of the thermopiles as a function of the applied temperature difference. Figures 7-a and -b report respectively the plain and PE thermopiles I-V curves for $\Delta T$ from 0 to 282 K temperature differences, respectively. Typical thermoelectric characteristics were obtained, and a generator regime (IV <0) was observed. The electrical current is given by

$$I = \frac{V_{G\,bias} - V_S(\Delta T)}{R_G} \qquad (3)$$

where $V_{Gbias}$ is the thermopile bias voltage, $V_S(\Delta T)$ is the Seebeck voltage at a given temperature difference across the thermopiles, and $R_G$ is the demonstrators' internal resistance, which is approximately 130k$\Omega$ and 260k$\Omega$ for plain and PE thermopiles, respectively. The electrical resistivity increased with PE, as confirmed by independent four-probe resistivity Van-der-Pauw measurements on dedicated areas of the wafer. A concomitant increase in the Seebeck coefficient and resistivity is expected from the Mott relation [28,29]. Because the dimensions of the PE lattice are much larger than the typical electron mean free path (a few nanometers), it is unlikely that the resistivity is significantly affected by the PE patterns. We attribute the origin of the resistivity increase to process fabrication-induced defects.

From the I-V curves, we calculated the output power corresponding to product I.V in the generation regime and extracted the maximum output power. Figures 7-c and -d present the maximum output power densities with respect to the temperature difference across the thermopiles (Figure 7-c) and according to the heating power of the Pt heaters (Figure 7-d). Figure 7-c shows that compared to the temperature difference across the thermopiles, the plain and PE demonstrators exhibit comparable performance. This observation reflects the compensation of the Seebeck effect enhanced by PE as the electrical resistance increased. The maximum output power is given by (4), and at a given temperature difference, it depends only on the Seebeck coefficient and electrical resistance.

$$P_{MAX} = \frac{V^2}{4 * R_G} = \frac{S^2 * \Delta T^2}{4 * R_G} \qquad (4)$$

where S and $\Delta T$ are the Seebeck coefficient and the temperature difference across the thermopiles, respectively. Globally, PE would therefore have no impact on generator performance. Figure 7-c also shows a comparison between the measurements and modeling studies from previous studies [23]. This second comparison also shows perfect agreement between the measurements and modeling results, validating the modeling studies.

The PE and plain demonstrators were compared again, but this time with respect to the heating power of the Pt heaters (Figure 7-d). Under these conditions, we can notice that, compared to the heating power, PE demonstrators exhibit better performance than plain thermopiles. This highlights the benefits of thermal gradient management for thermoelectric harvesting. Indeed, (5) gives the maximum output power, which depends not only on the Seebeck coefficient and electrical resistance that compensate each other, but also on the temperature difference across the thermopiles. The temperature difference across the PE thermopiles was higher for a given heating power (Figure 5).

$$P_{MAX} = \frac{V^2}{4 * R_G} = \frac{S^2 * Q^2 * r_{TEG}^2}{4 * R_G}$$

$$\Delta T = Q * r_{TEG}$$

(5)

Q and $r_{TEG}$ are the heat gradient across the thermopiles (or heating power) and the thermal resistance of the thermopiles, respectively. Figures 7-c and -d highlight then the fact that the main benefit of PE is better thermal gradient management, especially owing to a thermal conductivity reduction [13–15]. The gap between the PE and plain thermopiles in figure 7-d can be improved with a better PE design. The demonstrators exhibit hundreds of nW/cm$^2$ to a few mW/cm$^2$ of output power with respect to the temperature differences across the thermopiles. From this range of output power, we can easily imagine the use of a few cm$^2$ of such generators to supply power to autonomous sensor nodes [30]. Moreover, compared to the state-of-the-art (cf. figure 1), first with respect to the silicon-based thermoelectric microharvesters, the developed demonstrators mainly exhibit better performance than the state-of-the-art for the same temperature differences across the thermopiles. The material crystalline structure and/or harvester architectures can explain this difference in performance. Polysilicon has lower electrical resistivity than crystalline silicon. Low-dimensionality systems, such as silicon nanowires, feature improved thermoelectric efficiency. Tomita et al. [19] managed to develop Si nanowire (NWs)-based harvesters exhibiting state-of-the-art performance.

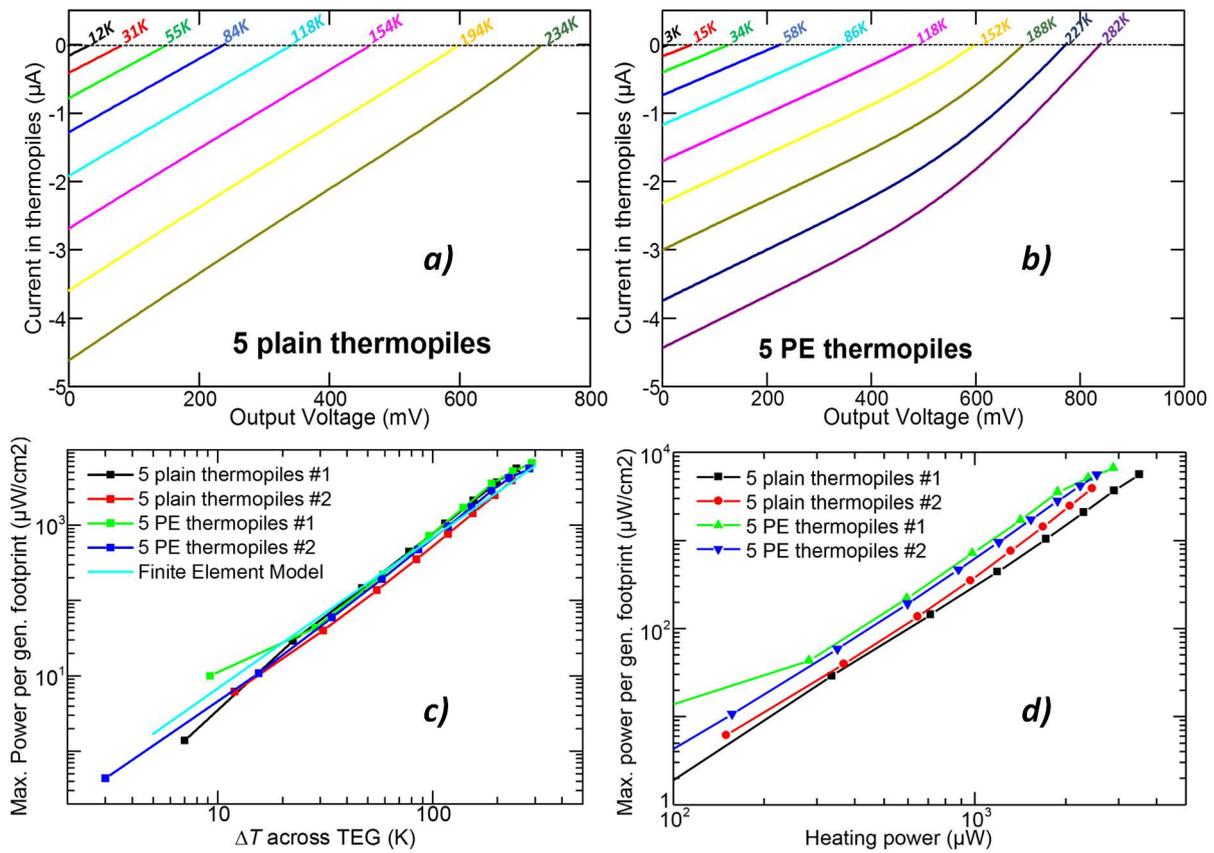

*Figure 7: 5 plain (a) and PE (b) thermopiles I(V) curves with temperature gradient and Measured maximum output power density versus temperature difference across thermopiles (c) and versus the Pt heaters' heating power (d)*

However, the performance of the vertical bismuth telluride micro harvesters was four orders of magnitude higher than that of the developed demonstrators. This is accounted for by the lower electrical resistance per thermopile for the vertical Bi-Te micro-harvesters, less than $1\Omega$ against tens of $k\Omega$ per thermopile for the developed demonstrators. Indeed, bulk (tenth of $\mu$m) $Bi_2Te_3$ exhibits low thermal conductivity, and there is no need for thinning, allowing for the development of harvesters with lower electrical resistances, in contrast to our demonstrators. Nevertheless, the modeling works presented in [23] demonstrated that according to the cooling conditions of the harvesters, the developed demonstrators could compete and even outperform the $Bi_2Te_3$ micro-harvester state-of-the-art (Figure 8).

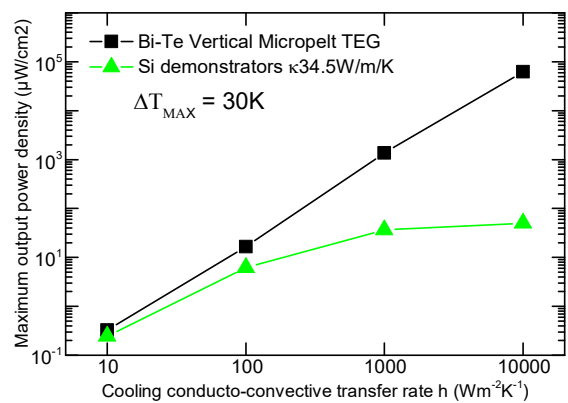

***Figure 8:*** *FEM modeling of a vertical Bi-Te TEG and planar Si demonstrator (34.5 W/m/K case) as a function of the cooling efficiency.*

Figure 8 compares the expected power density output as modelled by FEM for both the $Bi_2Te_3$ vertical TEG and planar Si TEG (34.5 W/m/K) under a maximal available temperature difference of 30 K. It is noticeable that a comparable performance is obtained under a weak cooling efficiency ($h < 100$ W/m²/K) [23]:

- Cooled with a low capacity of no heat sink, the performance of TEG is driven by the thermal gradient across the TE material, which is maximized in Si membranes.

- Cooled with a higher efficiency, the TEGs' performance is driven by the electrical resistance, which is lower for vertical $Bi_2Te_3$.

The characterization of the demonstrators is completed with the observation of Peltier cooling. Figure 9 shows infrared (IR) images of a demonstrator biased at ±7.5V. The left picture represents the result of current propagation from *p* to *n* end, and the right picture represents the current propagation from *n* to *p* end. The figure shows that, depending on the current direction, heat generation or absorption is observed at the center of the thermopiles, which is characteristic of the Peltier effect.

The heat generation and absorption were only observable at the center of the thermopiles because, first, the sample was maintained at 75°C during all the experiments, and second, the silicon membranes were too thin to absorb IR waves. The Peltier effect and, in particular, Peltier cooling, is then possible on silicon technologies, opening the way to a possible integration of thermoelectric coolers. However, it is worth noting that the TEG demonstrator was not designed for Peltier cooling, which is naturally much lower than Joule heating.

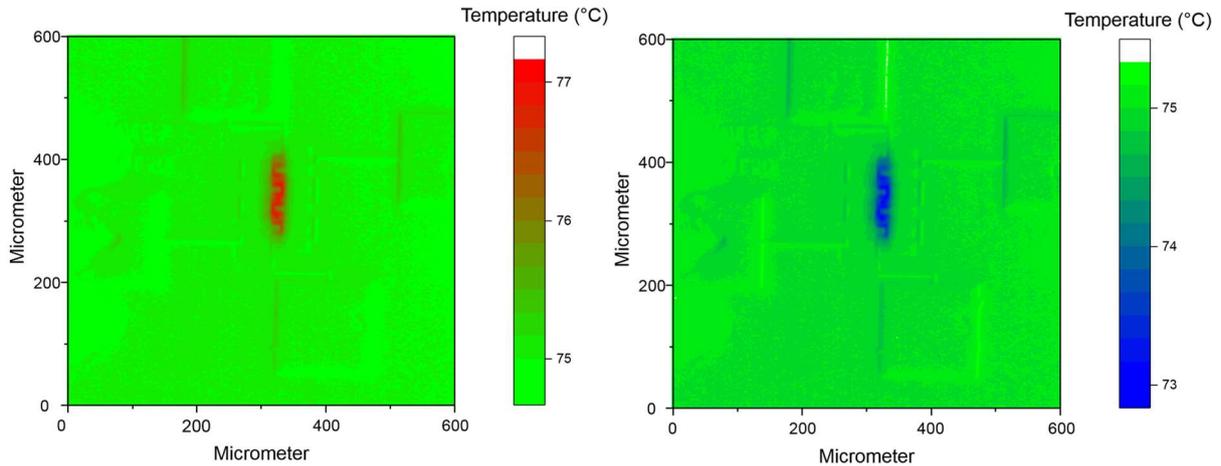

***Figure 9:*** *IR imaging pictures of the demonstrator with respect to the electrical current direction in the thermopiles. Left: current for p end to n end and right: current from n end to p end.*

## Conclusion

In summary, a phonon engineering approach was used to reduce the heat conduction in crystalline silicon thin membranes. The fabrication and demonstration of thermoelectric conversion using crystalline silicon was shown. The device principle relies on better use of the available thermal difference in a planar thin membrane architecture and additional phonon diffusive patterning. The results demonstrated the benefits of PE. Moreover, PE induces an increase in the Seebeck coefficient compensated by an increase in electrical resistance and better thermal gradient management through the thermopiles. The net gain remains positive and we report $400nW/cm^2$ under $\Delta T = 3$ K up to $6mW/cm^2$ under $\Delta T = 282$ K according to the temperature difference across the thermopiles. This CMOS-compatible converter opens new avenues for below-$cm^{-2}$ footprint energy harvesters for IoT power supply. In addition to the possibility to power supply autonomous sensor nodes, the studied demonstrators can be considered as state-of-the-art silicon-based micro-harvesters (Figure 1). Finally, the modeling study performed in [23] highlights the benefit of such micro-harvesters with respect to vertical bismuth telluride when they are cooled without any heat sink assistance (Figure 9). This study also highlights the feasibility of using integrated crystalline silicon thermoelectric coolers.

## Acknowledgments

This work has received: i) funding from STMicroelectronics-IEMN common laboratory; ii) funding from the European Research Council under the European Community's Seventh Framework Programme (FP7/2007-2013) ERC Grant Agreement no. 338179, iii) support from the French RENATECH network, iv) support from the NANO2017 program, and v) support from the French government through the National Research Agency (ANR) under program PIA EQUIPEX LEAF ANR-11-EQPX-0025."

## Data availability statement

All data that support the findings of this study are included within the article. More details regarding the fabrication process and characterization methodology can be found in the supplementary file.

## Competing interests

The authors declare the following competing interests: Patent US10103310B2